# Machine Learning and Computer Vision Techniques to Predict Thermal Properties of Particulate Composites

Fazlolah Mohaghegh, Jayathi Murthy

*University of California-Los Angeles; Department of Mechanical and Aerospace Engineering*

**Abstract**

Accurate thermal analysis of composites and porous media requires detailed characterization of local thermal properties in small scale. For some important applications such as lithium-ion batteries, changes in the properties during the operation makes the analysis even more challenging, necessitating a rapid characterization. We proposes a new method to characterize the thermal properties of particle composites based on actual micro-images. Our computer-vision-based approach constructs 3D images from stacks of 2D SEM images and then extracts several representative elemental volumes (REVs) from the reconstructed images at random places, which leads to having a range of geometrical features for different REVs. A deep learning algorithm is designed based on convolutional neural nets to take the shape of the geometry and result in the effective conductivity of the REV. The training of the network is performed in two methods: First, based on implementing a coarser grid that uses the average values of conductivities from the fine grid and the resulted effective conductivity from the DNS solution of the fine grid. The other method uses conductivity values on cross sections from each REV in different directions. The results of training based on averaging show that using a coarser grid in the network does not have a meaningful effect on the network error; however, it decreases the training time up to three orders of magnitude. We showed that one general network can make accurate predictions using different types of electrode images, representing the difference in the geometry and constituents. Moreover, training based on averaging is more accurate than training based on cross sections. The study of the robustness of implementing a machine learning technique in predicting the thermal percolation shows the prediction error is almost half of the error from predictions based on the volume fraction.

## 1- Introduction

Particulate composites consist of little particles of one substance embedded in another substance to improve material properties. For example, in tire manufacturing carbon black particles are bonded with a rubber matrix to form an abrasion resistance material with enhanced tensile strength [1]. Brake disks made of a carbon-ceramic composite are 65% lighter than the traditional cast iron ones and maintain a high friction coefficient even at elevated temperatures [2]. In electronic packaging, the enhancement of polymers with metal particles leads to a composite with polymer mechanical properties and boosted thermal or electrical conductivities [3]. In general the properties of the particulate composites depend on the distribution, shape, and size of filler particles [4]. Due to the high number of effective parameters, the study of these systems is mostly limited to experiments or models with many simplifications. This work proposes a new approach of studying the effects of particle geometry on thermal conductivity of particle composites focusing on lithium-ion battery electrodes.

Lithium-ion batteries are known as one the most popular rechargeable batteries that are widely used in portable electronic devices due to low weight, high energy density [5], low memory effect [6], and little self-discharge[7]. The batteries electrodes including rechargeable batteries are made of composites of different materials with different geometrical features, making their analysis challenging. The rechargeability of the battery stems from the fact that lithium ions can move from the intercalated compound of the cathode to the anode during the charge process [8]. The presence of lithium ions in the domain is in the form of active cathode particles like $LiMn_2O_4$, $LiCoO_2$ and $LiFeO_4$ which are dispersed in a binder such as PVDF (polyvinylidene fluoride) imbedded with carbon particles to increase electric conductivity. This particulate composite forms a porous medium wherein an electrolyte fills the pores and wets the active particles. The performance of a battery hinges on how much the electrode structure facilitates the transport of lithium ions in contact with the electrolyte.

While widely used, lithium-ion batteries still suffer from a few fundamental problems [9] such as thermal issues, which are considered as one of the critical problems restricting their reliability, applicability, and safety [10-12]. Thermal runaway occurs when insufficient heat dissipation leads to temperatures above the critical limits, about 80℃ [13], and exothermic reactions start a positive feedback loop and consequently fire or explosion [14]. In addition to irreversibilities, the recoverable power and the capacity of the battery reduces significantly at temperatures above 50℃ [10]. A proper parametric study to optimize the thermal or electrical performances is a cumbersome task because of a high number of influencing properties and the complexity of the nonlinear system [9]. The impact of different parameters such as volume fraction of materials or the size of active particles on the battery performance has been often studied in experiments extensively [15-19]. However, if the numerical modeling is established, the parametric optimization of the battery becomes significantly easier.

To study thermal issues, first the thermal properties e.g. heat capacity and effective thermal conductivity of the battery components should be characterized properly. The numerical study of heat conduction in a heterogeneous media such as battery electrodes involves combining the heat transfer for different phases. The volume averaging technique is widely used to reach computationally less expensive calculations [20-25]. The volume averaging method by defining the average properties over a volume element and in a multiphase system obtains solvable equations to predict macroscopic properties [26]. However, the volume element loses the individual phases' information if the volume contains more than one phase [26]. In addition, the accuracy of volume averaging models depends on how properly the closure terms are defined [27].

The alternative to the volume averaging method is the fully resolved direct numerical simulation (DNS) study in the small scale to include the details of pores and particles geometries [28]. The initial DNS studies were implementing the geometry involving simplifications such as use of homogenized values from porosity or tortuosity to describe the entire electrode's effective parameters [29, 30]. Other studies added the inhomogeneities in the form of arrays of particles with canonical shapes e.g. spheres or ellipsoids or a complex of them [31-33]. Obviously, these simplifications lead to deviation from reality as the active particles' topology and their contact places cannot be defined accurately based on canonical shapes. A better understanding could be achieved by 3D simulation on real active particle shapes. The particles have dimensions with micro-scale size and have irregular shapes. DNS can properly use the real 3D micro-images of electrodes' structures and thus, include the geometrical details and the topologies of different constituents. Fortunately, nowadays with the advent of instruments like beam-scanning electron microscopes (FIB-SEM) [34] and micro-computed tomography ($\mu$CT) or (SRXTM) [35] the micro-images of the electrode are available. Using the actual particle images to generate the 3D meso-scale simulation geometry of pores and particles has been reported in [9, 17, 32, 36, 37].

While implementation of actual electrode images significantly increases the robustness of the study of heat transfer in Lithium-ion batteries, the real challenge still exists as the electrodes shrink or swell due to insertion and de-insertion of lithium during the charge/discharge processes. In addition, the imposed stresses during the cycling leads to development of cracks. Therefore, the volume fraction of the constituents, particles size, topology of electrolyte and electrode interface, porosity of the active particles, and geometrical shape of the electrode are dynamically changing [38-40]. Each of the mentioned parameters affects the transport of charges and heat in the battery as they indicate the contact surface of the electrodes with the electrolyte and define the material properties of cell components. Optimization of battery performance with several number of effective parameters necessitates very efficient simulation methods which can find the final performance of the system in low cost.

Recently, the machine learning techniques have gained popularity in engineering applications such as image processing [41], renewable energy [42], robotics [43] and software engineering [44]. Machine learning employs statistical methods to train a computer system with available data to improve the performance of a computer program through experience. Therefore, when a machine learning algorithm is trained, the computational cost to solve a problem significantly decreases. We believe that machine learning is a good candidate to characterize the heat transfer in a lithium-ion battery electrode. This paper proposes a new approach to predict a battery's thermal properties based on machine learning techniques. The method takes the real 2D stack of electrode images as the input and then constructs the 3D computational grid from the stack of images. Performing DNS simulations on random samples from the grid provides the training data set. Then with using either averaging or cross section the network is trained through a convolutional neural network (CNN). The result show that the saved trained network can predict the thermal behavior of the electrode with 3.5% accuracy and at significantly shorter time. Therefore, the method can be used to optimize a battery heat conductivity when having several affecting parameters necessitates running numerous cases.

## 2- Theory and Modeling

The following energy conservation equation represents the intrinsic heat conduction equation in porous media [45]:

$$\frac{\partial \left(\rho_i C_{p_i} T_i\right)}{\partial t} = -\nabla(k_i \nabla T_i) + Q_i \tag{1}$$

where $\rho_i$, $C_{p_i}$, $T_i$, $k_i$, and $Q_i$ are respectively the density, heat capacity, temperature, heat conductivity, and total heat generation in the bulk of phase $i$. The goal is to find the effective thermal conductivity, $k_{eff}$, of the electrode from the images in the most effective way. In this study, the heat generation is neglected due to low electric resistivity in electrodes. Also, because changes in temperature occur very slowly, we consider a steady state situation and neglect the transient terms. Thus, the equation (1) becomes:

$$\nabla(k_i \nabla T_i) = 0 \tag{2}$$

To reconstruct the geometry, we use stacks of 2D micro-images available from different studies to build a 3D grid based on the actual electrode geometry. We use images from Material and Device Engineering Group at ETH [46], which are obtained by X-ray tomography (SRXTM) [47]. SRXTM generates a stack of 2D image slices, which form a full 3D microstructure when they are stitched together. All samples are in discharge situations before cycling.

SRXTM is a non-destructive technique that uses X-rays to generate cross sections of the physical object in order to construct a 3D model [48]. It works based on differences in the absorption and scattering in the source of X-ray to reveal the internal physical attributes of the object structure. The method is used to characterize the microstructure of battery components [49], porous media [50], and medical scanning [51]. The first group of images in this study, represent a cathode made from NMC (LiNi$_{1/3}$Mn$_{1/3}$Co$_{1/3}$O$_2$) with PVDF binder and carbon black. Fifteen different electrodes are used with different compressions i.e. 0, 300, 600 and 2000 bars and various percentile weight of carbon black and binder i.e. 2-5 wt% for each. The NMC particles in all 15 electrodes have almost spherical shapes with approximately $5 - 15 \mu m$ diameters. The 3D microstructure in this study has a size of $700 \times 700 \times 70 \mu m^3$. Each image in the stack of 2D images has $889 \times 889$ pixels representing a layer from the 3D image. The other group of images from ETH group belongs to X-ray tomography of four different commercial graphite anodes. The samples have a diameter of 700 µ$m$ and thicknesses of 48.75 $µm$, 68.25 $µm$, 55.25 $µm$, and 74.75 $µm$. The amount of binder and additives are limited to 2-4 wt%. The size of the individual graphite particles ranges from 5– 30 $µm$ with flake-like shapes in electrode I, III, and IV and spherical shape for electrode II. Figure (1) shows sample of the 2D images implemented in this study.

All image stacks are already preprocessed to transform from grayscale to a black and white binarized scale. The binarization process includes filters to reduce the noises and clarify the boundaries of phases in the images. The white areas correspond to the lithium as the active material and the black regions belong to the non-active areas i.e. pores, carbon black and binder.

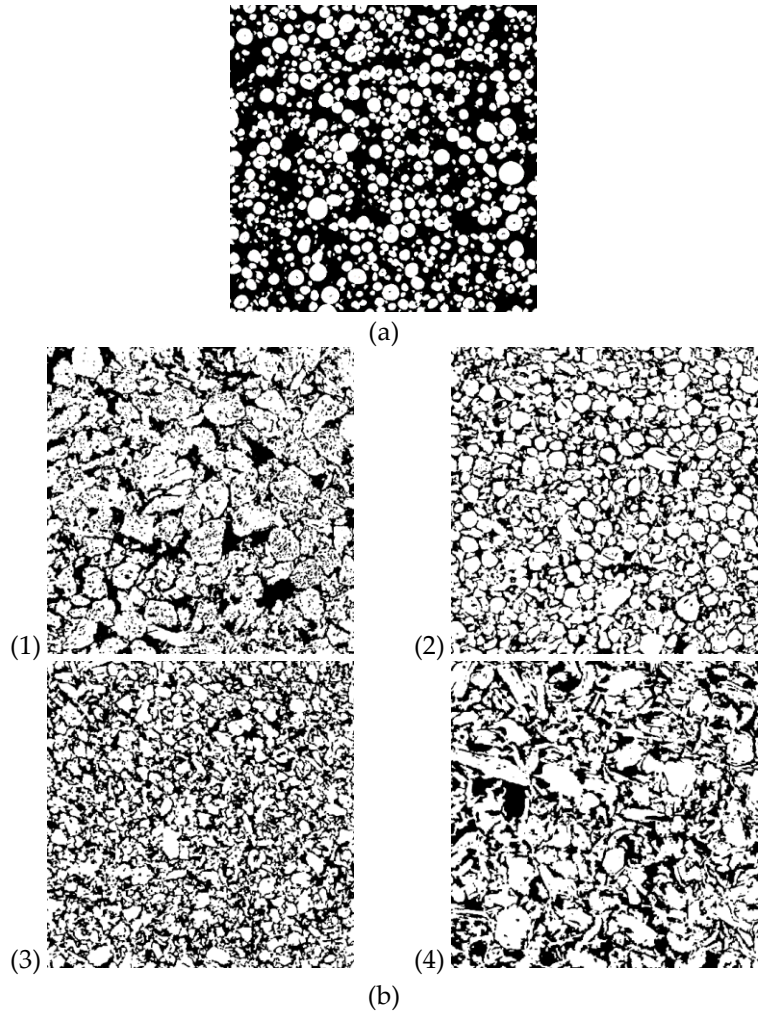

*Figure 1- Samples of 2D images used in this study: (a) NMC electrode [47], (b) graphite electrodes [52].*

In DNS, the explicit partitioning into different constituent phases at pore-scale employs a length-scale, which is several orders of magnitude smaller than the macro-scale making the simulations computationally intractable. A representative elemental volume (REV) can bridge the gap between the scales by transferring the data from small-scale to large-scale. REV is a spatial volume with small dimensions compared to the cell size but large enough to contain enough particles to statistically capture the essential geometrical and topological information. A fine grid resolution simulation at pore-scale finds the effective conductivity of REV containing the detailed geometries of particles and pores. Then, the effective conductivity of REV is used in the macro scale simulations by considering a single-phase material for the battery electrode with properties obtained from REV. Figure (2) shows a schematic of the 3D constructed images that describes how REV is chosen in a domain. For simplicity we choose a cube shape for the REV and perform the computations on REVs at different random locations. Then the effective thermal conductivity of REVs could be used in macro-scale to study the heat transfer at cell scale.

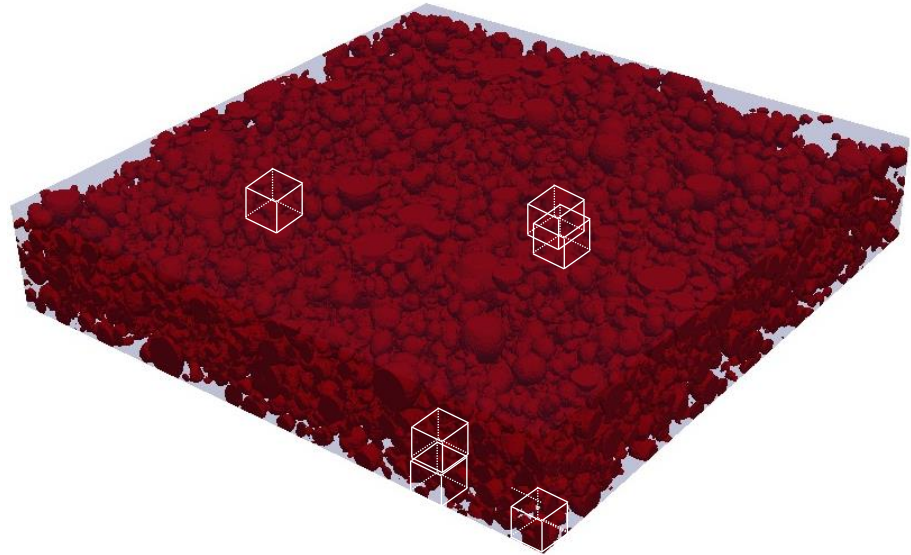

(a)

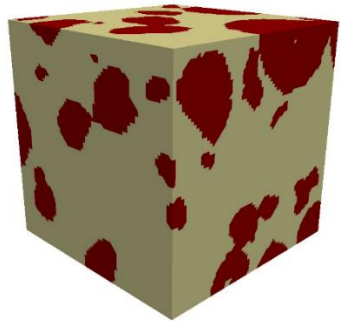 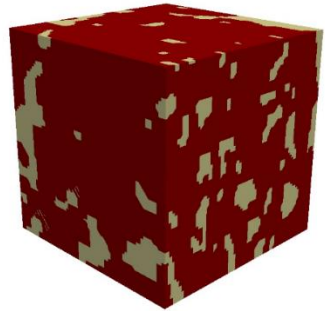

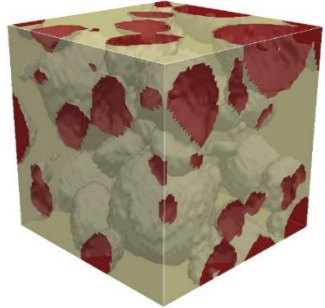 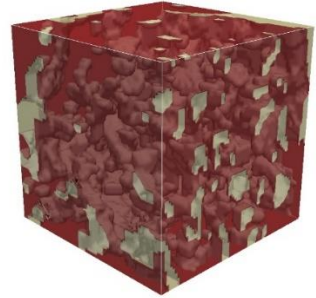

(b)         (c)

*Figure 2- (a) A schematic of the reconstructed geometry and REV cut in the 3D domain for NMC electrode [47]. REVs obtained from the constructed 3D image for (b) NMC electrode [47], (c) graphite electrodes [52].*

The image analysis is performed using the Python Image library, which can import an image, extract the RGB values and crop the image. To generate 3D REV images, after random selection of the starting point at the REV corner, the images are trimmed in $x$ and $y$ direction in 2D images and then consecutive images in the stack are chosen using the same trimming location and same $x$ and $y$ directions. The volume fraction is defined by counting the ratio of the total number of white pixels to the total number of pixels in the REV i.e. the volumetric fraction of the active lithium particles in the electrode. Because the binder and the additives have a small fraction of the electrode weight, their conductivities are considered in conjunction with the electrolyte as $k_2$. In figure (1), black pixels represent $k_2$ and white pixels represent $k_1$. Different electrolyte materials and varying weight fractions of the constituent means that the normalized value of the conductivity i.e. the conductivity ratio $k_2/k_1$ also varies significantly. Therefore, a range of $k = k_2/k_1$ should be considered to perform the analysis.

To find the effective thermal conductivity, the boundary condition setup chooses two different arbitrary temperatures $T_h$ and $T_l$ at two opposing sites. The rest of the boundaries are considered as adiabatic. Because the temperature range in a battery is not significant, the assumption of constant uniform heat conductivity is valid. Solving equation (2) results in finding the temperature distribution in the domain. The effective thermal conductivity in direction $x_m$ is computed after calculating the heat rate $q$ by integration over a plane perpendicular to direction $x_m$:

$$k_{eff,x_m} = \frac{q}{\frac{\partial T}{\partial x_m} A} \tag{2}$$

Each side of the cubical REV has a size of 56.62 $\mu m$ leading to a $72 \times 72 \times 72$ grid. Each pixel in the image is assigned with a grid cell. The finite volume solver uses BiCG sparse linear solver to calculate the temperature distribution.

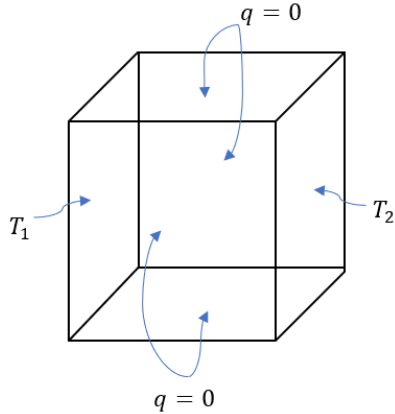

Figure 3- The boundary conditions for the study of effective conductivity.

## 3- Machine Learning Algorithm

In machine learning, the training means to describe the network parameters i.e. weights and biases by using the available data so that the network can make predictions without the necessity of solving the original relations. Therefore, the input to the network is a matrix representing the distribution of different material conductivities in the REV and the output is the effective conductivity of REV. Because the input data is big, we use convolutional neural nets (CNN) to construct the network. CNNs are known to work better on large data than the artificial neural nets (ANNs) by lowering the errors and performing faster [53]. A CNN, instead of using regular matrix multiplication, uses the concept of convolution as the hidden layer of a network. A convolutional layer is formed by sweeping a set of kernels over the image to form a
6

filter and extract the main features through a pooling [54]. While we implement CNNs to cope with large data, the training based on the REV with the size of 72 × 72 × 72 is still a tedious task. When the input data is very big, the network should be deeper, the training becomes slow, and most of the time the final error in training is high. Moreover, the memory limitations restrict the size of the network and the size of the input data. Here, two approaches are suggested to overcome these difficulties. The first approach is based on averaging the values of thermal conductivities related to each cell and the other approach is based on choosing the thermal conductivity on a cross sections in each REV.

*3-1- Training Based on Averaging*

In this approach, as shown in Figure (4) instead of using the original fine grid as the network input, a coarser grid is implemented by averaging the values of heat conductivities from the fine grid. With coarsening, the conductivity matrix becomes smaller and the training will be faster and with smaller error. The averaging introduces new conductivity values between $k_1$ and $k_2$, which occurs at the interface of the two materials i.e. at the interface of the black and white regions in Figure (1). As shown in Figure (5), the coarser the grid is, the more smeared the interface will be. To apply the averaging, each cell in a coarse grid will contain a cube of cells from fine grid but with conductivity values obtained from the averaging over the fine grid cells. We will study the effect of coarse grids on predicting the values of effective conductivity in the next section. Table (1) shows the different grid sizes and their related neural nets used in this study.

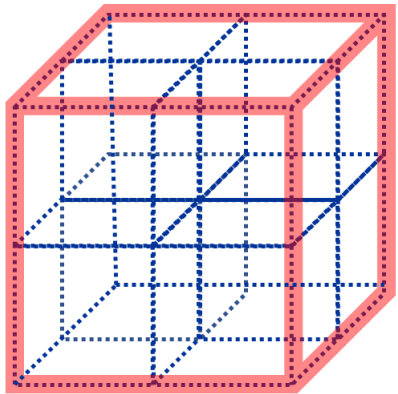

*Figure 4- A grid coarsened by 2 (red grid) used in the training process.*

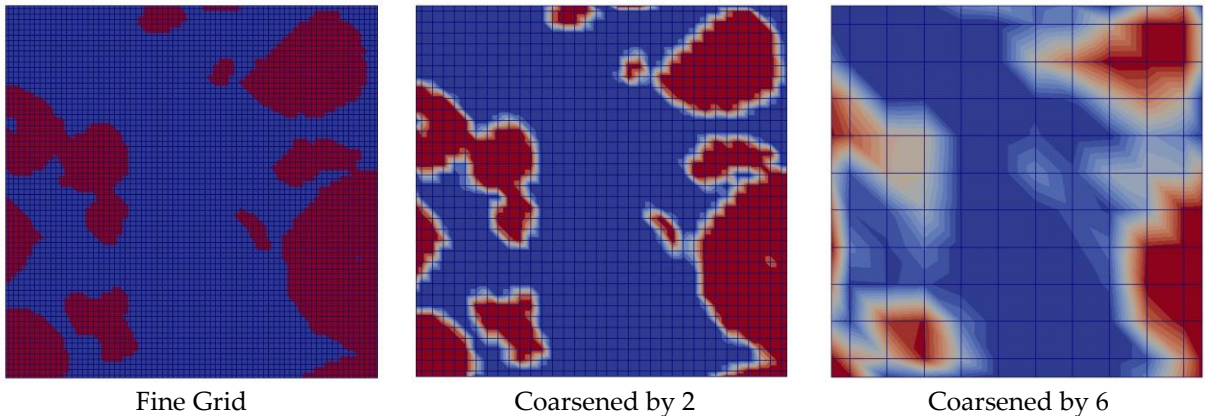

Fine Grid          Coarsened by 2          Coarsened by 6

*Figure 5- Grids with different levels of coarsening used as the input to each machine learning network.*

Table 1- Size definition at different places of the network for training based on averaging

| Coarsening Level | Grid | Input | CNN1 | Max pool 1 | CNN2 | Max pool 2 | FCC 1 | FCC 2 |
|---|---|---|---|---|---|---|---|---|
| 1* | 72 × 72 × 72 | 864 × 432 | 861 × 429 × 32 | 287 × 143 × 32 | 248 × 140 × 64 | 94 × 46 × 64 | 7 | 7 |
| 2 | 36 × 36 × 36 | 216 × 216 | 213 × 213 × 32 | 71 × 71 × 32 | 68 × 68 × 64 | 34 × 34 × 64 | 7 | 7 |
| 3 | 24 × 24 × 24 | 144 × 96 | 141 × 93 × 32 | 47 × 31 × 32 | 44 × 28 × 64 | 22 × 14 × 64 | 7 | 7 |
| 4 | 18 × 18 × 18 | 108 × 54 | 105 × 51 × 32 | 35 × 17 × 32 | 32 × 14 × 64 | 16 × 7 × 64 | 7 | 7 |
| 6 | 12 × 12 × 12 | 48 × 36 | 45 × 33 × 32 | 15 × 11 × 32 | 12 × 8 × 64 | 6 × 4 × 64 | 7 | 7 |
| 12 | 6 × 6 × 6 | 18 × 12 | 15 × 9 × 32 | 7 × 4 × 32 | 7 × 4 × 64 | 4 × 2 × 64 | 7 | 7 |

*The original fine grid has these extra layers: CNN3 with size 91 × 43 × 128, and max pool3 with size 30 × 14 × 128

Except for the fine grid network, which has three layers of convolutional layers, the rest of the networks have two convolutional layers. The convolutional layers provide the input for two layers of affine layer layers i.e. fully connected layers. For the first convolutional layer, 32 4 × 4 filters are implemented over the reshaped input array. Then, a max pooling layer with a 3 × 3 kernel and (3,3) strides selects from the output of the convolutional layer. The second convolutional layer has 64 filters and feeds a similar max pooling layer but with a 2 × 2 kernel and (2,2) strides. The output is reshaped to feed into two fully connected layers, each with seven nodes. A leaky-ReLU activation function applies the non-linearity to the network in each of the mentioned layers. Figure (6) shows a schematic of a machine learning algorithm for a 36 × 36 × 36 grid. The concept is not different for different grids as only the number of layers and their sizes change with the change of the input data.

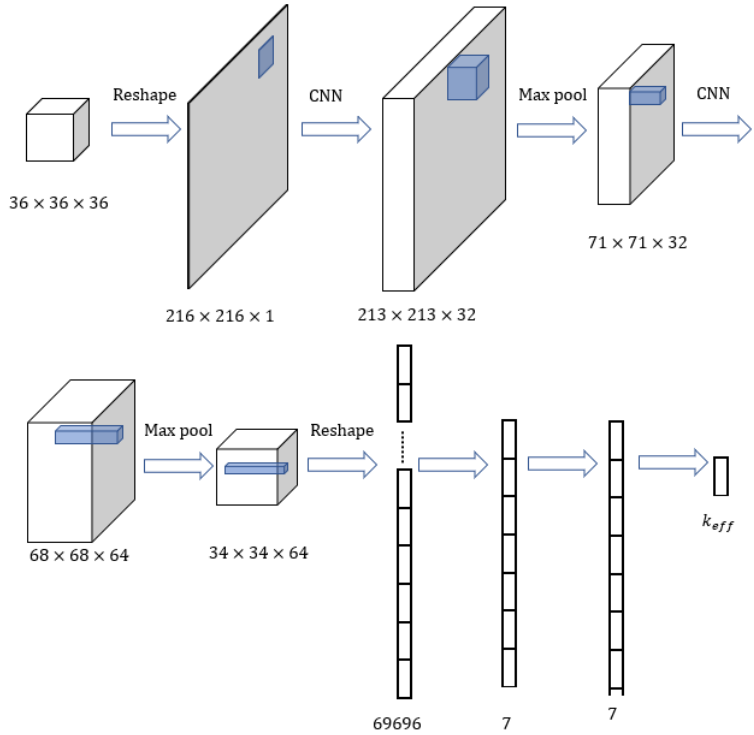

Figure 6- The structure of the network used to predict the effective conductivity in an REV.

The training has been accomplished on each network for the number of 50,000 training steps based on random places of REV and random values of $k_2/k_1$. In each training step, Adams optimization [55] with the learning rate of $10^{-4}$ performs a stochastic gradient-based optimization on mini batch sizes of 50.

### 3-2- Training Based on Cross Section

This method chooses up to three cross-sections in the middle of the REV in different directions: $x$, $y$ and $z$. It uses the values of thermal conductivity on each cross section with the grid size $72 \times 72$. These conductivity values as well as the results of effective conductivity from fine grid DNS solution are used as an input to a network similar to the ones in section (3-1) but with the sizes shown in Table (2). In this case, the learning rate is 0.00002.

*Table 2- Size definition at different places of the network for training based on cross section*

| Coarsening Level | Grid | Input | CNN1 | Max pool 1 | CNN2 | Max pool 2 | FCC 1 | FCC 2 |
|---|---|---|---|---|---|---|---|---|
| 1 | $1 \times 72 \times 72$ | $72 \times 72$ | $69 \times 69 \times 32$ | $23 \times 23 \times 32$ | $20 \times 20 \times 64$ | $6 \times 6 \times 64$ | 7 | 7 |
| 2 | $2 \times 72 \times 72$ | $144 \times 72$ | $141 \times 69 \times 32$ | $47 \times 23 \times 32$ | $44 \times 20 \times 64$ | $14 \times 6 \times 64$ | 7 | 7 |
| 3 | $3 \times 72 \times 72$ | $216 \times 72$ | $213 \times 69 \times 32$ | $71 \times 23 \times 32$ | $68 \times 20 \times 64$ | $22 \times 6 \times 64$ | 7 | 7 |

## 4- Results

We first validate the DNS solver by comparing DND results with a benchmark. Then, we study the average-based training and investigate the affecting parameters. In the next step, we study the cross-section-based training and compare the two training methods. Finally, we study the capability of the neural nets in the study of thermal percolation.

### 4-1- Validation of The DNS Solver

For the validation of CFD simulations, we compare the predictions of effective conductivity from DNS with the results of Vadakkepatt et al. [56], obtained from the volume averaging technique for NMC cathode. Figure (7) shows the overall agreement between the two methods. The results of Vadakkepatt et al. [56] for the effective thermal conductivity are the averaged values based on the volume fraction of the lithium particles, representing a range of results for each volume fraction. The current results also show a scattered data showing that for each conductivity ratio, the volume fraction is not the only parameter affecting the thermal conductivity. In fact, the particle shape and the geometry also play a role, especially when the conductivity ratio increases. At high conductivity ratios, the effective conductivity has more scattered results due to more dependency to the geometry.

For further clarification, we study the effect of geometry by studying different types of graphite anode images. Random REVs are generated here and REVs with a volume fractions, $vf \sim 0.65$ i.e. within the range of $0.64 < vf < 0.66$ are selected for the DNS study. Here, $k_{eff}$ is calculated for $k_2/k_1 = 50$. Because the value of $k_{eff}$ varies with the place of REV for each type of graphite, a total of 40 REVs are chosen and their averaged value of $k_{eff}$ are used. It is clear from Table (3) that even though the volume fraction of the constituents is the same for all graphite types, the mean value of $k_{eff}$ varies with different standard deviations. This shows the dependency of the conductivity on the geometrical features of electrode.

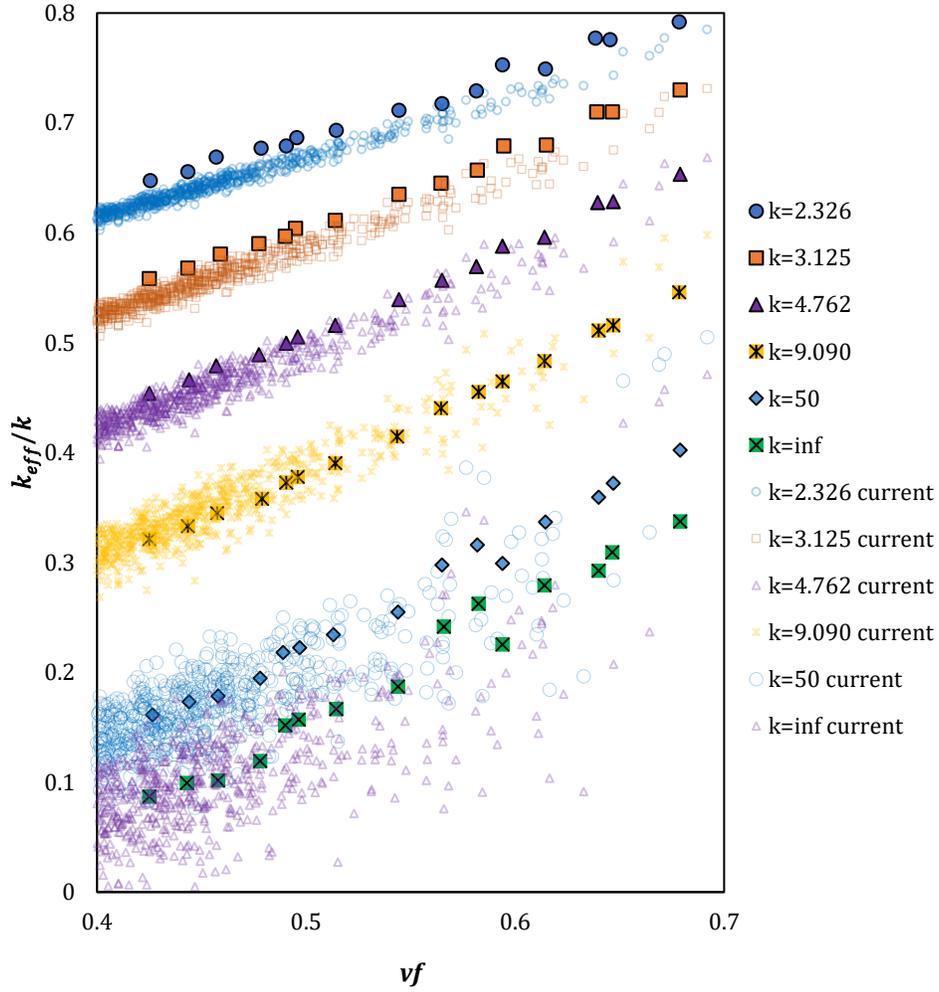

Figure 7- Effect of volume fraction, $vf$, on the normalized effective conductivity, $k_{eff}/k$, at different conductivity ratios, $k = k_2/k_1$.

Table 3- Variation of the average effective conductivity and standard deviation for graphite electrodes with various geometries.

| Type | Mean Effective Conductivity | Standard deviation |
|---|---|---|
| 1 | 21.348 | 2.371 |
| 2 | 18.156 | 1.812 |
| 3 | 20.025 | 1.436 |
| 4 | 24.260 | 1.433 |

### *4-2- Network Training for The Cathode*

We start with the training based on averaging. The training is performed over 2000 training and 200 testing samples based on random places of REVs in the 3D reconstructed geometry. The error in the network is obtained by using the trained network to predict the effective conductivities in the testing samples and then using the average of the errors on the total number of training samples. For each REV, we choose a random value between 0 and 10 for the conductivity ratio. The training is performed on a network containing images of an NMC cathode, with manufacturing pressure of 0 bar and weight fractions of 5% for carbon black and PVDF. The input to the network related to the original fine grid is the matrix of conductivities with size of $72 \times 72 \times 72$ and element values of either one or $k_2/k_1$. For coarser grids, the matrix size shrinks but averaging leads to introducing other matrix element values between one and $k_2/k_1$. During the training, matrix of conductivities on each grid point is used as the input to the network. The value of effective conductivity from the fine grid DNS solution is used to compare the generated output and update the network accordingly. Therefore, even for the coarse grid data as the input to the network, still the effective conductivity values from the fine grid solution is used.

Figure (8) shows the effect of using averaging on the prediction error and on the training time. Clearly, different levels of coarsening do not have a significant effect on the network error. Using the effective conductivity from the fine grid DNS solution helps to preserve the network accuracy on fine grids. When a coarser grid is used, the updated input data loses its originality and the network error should increase. However, the increase is balanced with the better training for the smaller number of data due to having smaller input matrix. Figure (8) also shows the effectiveness of this approach in reducing the training time. Use of the coarser grid instead of a fine grid significantly changes the training time from about a week to about 9 minutes. This improvement in training efficiency does not notably change the network error.

Now, the trained network is used to predict different types of NMC cathodes. All NMC cathodes have close to spherical particle shapes but they are different based on different weight fractions of the constituent and the manufacturing pressure i.e. rolling pressure. The weight fractions of carbon black and PVDF changes equally in each electrode. Figure (9) shows the effect of using a trained network which was coarsened by 2 in order to predict the effective conductivities over a set of 100 testing samples. The network is accurate in predicting other types of cathode as the particles' shapes are not changing meaningfully.

One interesting benefit of this approach is the significant computational cost reduction. We are using an intel Core-i7-6700K CPU @4.00 GHz with 32 GB of RAM. For a trained network, it takes about 18 s to read the data from the trained network, read 100 images and calculate the effective conductivities. The related computation time for the DNS solver is about 1310s showing that the machine learning approach is about two orders of magnitude faster than the DNS. This significant difference in the efficiency suggests implementation of a machine learning algorithm when the solution should be performed over analogous data.

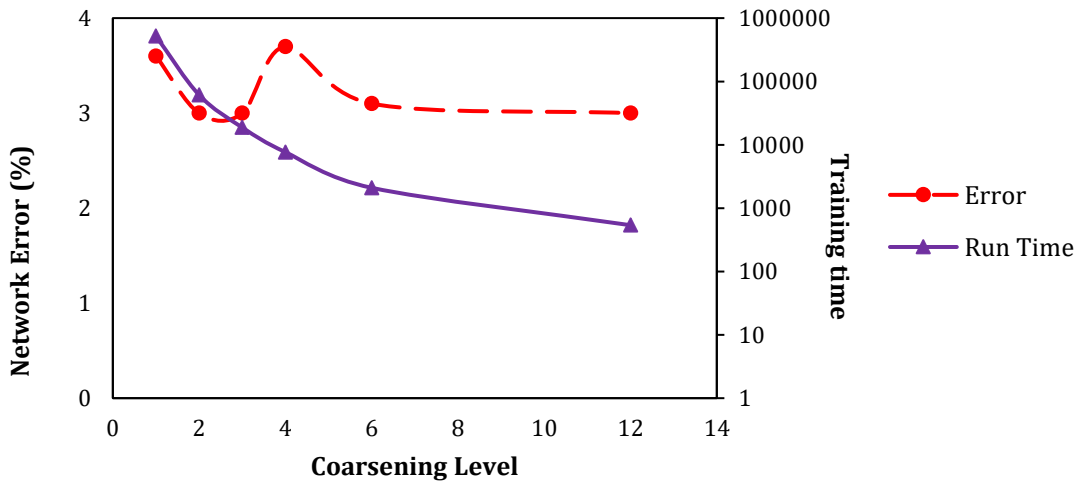

*Figure 8- Effect of coarsening level on the prediction of effective thermal conductivity and the running time.*

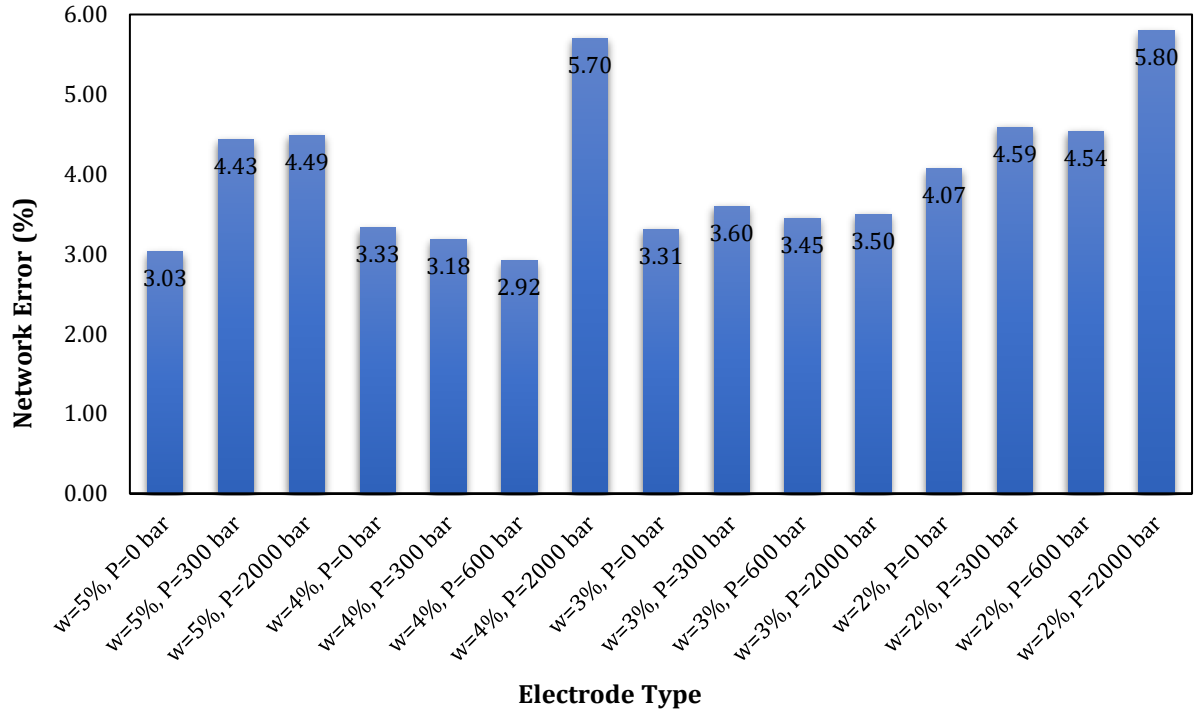

Figure 9- Effective thermal conductivity versus NMC electrode type using the trained network.

### 4-3- A Network for Both Cathode and Anode

As the cathode and anode have different textures, use of a network solely trained based on the cathode for the prediction of thermal properties in anode leads to high errors. Thus, to have a more general network, adding other types of electrode images is necessary. Here, to reach a more comprehensive network, we generate 22000 training samples, 10000 of which are based on the cathode images, and 12000 based on the four different sets of anode images. Therefore, the number of training samples is about one order of magnitude greater than the training samples in section 4-2. The number of testing samples in cross validation is 10% of the training samples. We use the same network size as before (Table 1) but this time, the training is based on a greater number of input samples. With a coarsening level of 6, the results show an error as small as 2.0% for all electrodes for $0 < k_2/k_1 < 10$. This reduction occurs in the error occurs due to using a greater number of training samples.

### 4-4- Effect of Conductivity Ratio

As described in section (4-1), when $k_2/k_1$ increases, the dependency on the geometry also increases, leading to more variation of effective thermal conductivity versus volume fraction. In other words, for the same volume fraction of NMC, a wider range of effective conductivity is expected. This issue is investigated by changing the range of $k_2/k_1$. Also, the results obtained from training of the network based on the geometry are compared with the results obtained from a network, which solely is trained with the volume fraction and $k_2/k_1$. The latter network is an artificial neural net with three hidden layers with 8, 40, and 8 nodes in the consecutive layers. A Leaky_ReLU activation function and learning rate of 0.001 is used in the network. The training and testing sizes of the two networks are the same as section (4-3). The results (Figure

10) show that in both networks, when the range of $k_2/k_1$ increases, the error also increases because the data is more sensitive to the training. However, the error of the network, when the training is based on the geometry is about 40% less than the error when the training is based on the volume fraction and $k_2/k_1$.

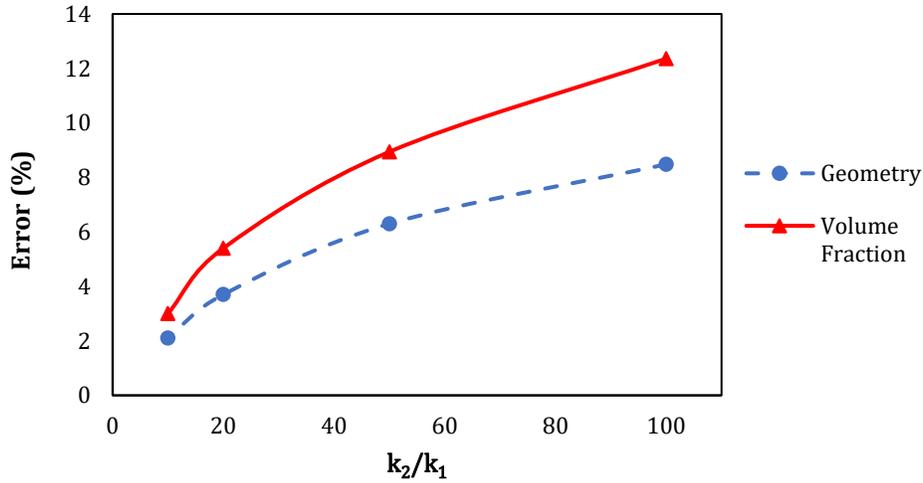

*Figure 10- Effect of thermal conductivity ratio on the network error: With increase of $k_2/k_1$ the network error increases. Training based on the current approach is more accurate than training based on vf and $k_2/k_1$.*

### 4-5- Training Based on Cross Section

In this section, we use up to three cross sections in each REV at different directions to perform the training. The chosen cross sections are in the middle of the sides perpendicular to the principal direction and they represent some of the geometrical features. The values of effective conductivity are achieved from DNS results.

As Figure (11) shows, with increase of the number of cross sections, the error decreases. This is because more cross sections, holds more geometrical features, leading to less error. However, the error in comparison with training based on averaging is noticeably higher. This is because the cross section is not a good representation of geometrical features. Considering the REV for NMC in Figure (2), if the place of the cross-section plane is changed across the $x$ direction, then the volume fraction changes from 0.33 to 0.59 showing a significant range.

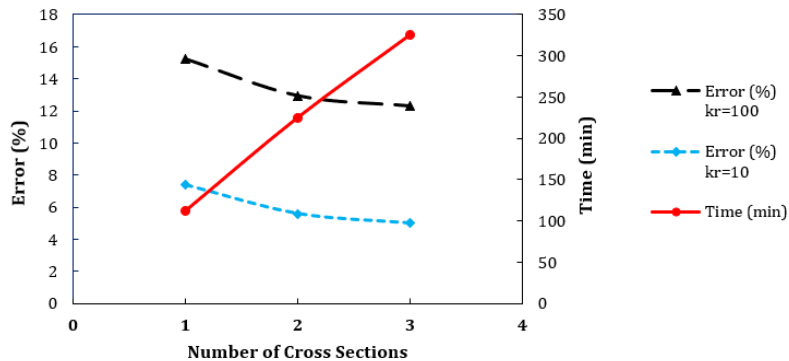

*Figure 11- Training based on the cross section: With increase of number of cross sections the network error decreases. However, the error is greater than the training based on the averaging.*

*4-6- Percolation Analysis*

Percolation is the passage of an effect over an irregularly shaped medium wherein that effect can more easily pass over some regions than others [57-59]. Thermal percolation studies the heat rate through random-shape composite e.g. a battery electrode as porous medium. When $k_2/k_1 \gg 1$, depending on volume fraction and the distribution of filling material, the heat conduction is almost negligible in some regions while being very high in other regions. When the volume fraction is more than a threshold, depending on the geometry heat passes may form in the medium [60]. Therefore, with the increase of volume fraction, a smooth and fairly sharp increasing trend [61] in the effective thermal conductivity occurs. This non-linear behavior shows the dependency on heat passes i.e. the geometry of the embedded structure in addition to the volume fraction.

In this part we investigate the applicability of machine learning techniques in percolation analysis. We generate 3D geometries with grid sizes $72 \times 72 \times 72$, each containing either cubic particles with the size $5 \times 5 \times 5$, or elongated particles i.e. rods with the size $5 \times 5 \times 30$. For a domain containing cubes, the number of particles is a random value between 1 and 550, and for the domain containing rods, the number of particles is a random value between 1 and 320 leading to volume fractions between almost zero to almost 0.36. No overlap is allowed but the particles can touch each other and form a heat pass. The particles are placed in the domain randomly with the following algorithm: First, we decide on the type of the particles. All particles have the same type in each domain. Then, we assign the value of $k_1$ as the conductivity for all nodes in the domain. Afterwards, a particle is added to the domain with random values of the position and orientation. The values of the conductivity in places that the rod exists change from $k_1$ to $k_2$. To add a new particle to the domain, first we assign a random position and a random orientation to a new particle. Then, we check to find if there is any node with the conductivity $k_2$ in the considered place of the particle. If there is any node with conductivity $k_2$, new random values for the position and direction are generated and assigned for the particle. If not, the particle is added to the domain by changing the conductivity values related to the nodes inside the considered place from $k_1$ to $k_2$. Figure (12) shows samples of the generated geometry this way.

Figure (13) shows the variation of effective thermal conductivity with the volume fraction for a porous medium with $k_1 = 1$ and $k_2 = 1000$. This figure represents DNS results of 5000 random non-overlap cubic and 5000 similar but elongated particles. It shows the non-linear independency of the volume fraction and a wide range of $k_{eff}$ for $vf > 0.15$. Different polynomial regression analyses were performed and the third order polynomial $k_{eff} = 521.86 \, vf^3 - 94.032 vf^2 + 11.321 vf^2 + 0.9688$ with $R^2 = 0.697$ is selected as the best fit. The fair value of $R^2$ signifies the importance of the geometry of the embedded material in addition to the volume fraction. The error in the regression prediction based on volume fraction is 30.4%. From Figure (13), the dependency on the geometry increases with the increase of the volume fraction because more heat passages may form randomly with an increase in the number of rods. However, because of the randomness, the resulted figure represents a scattered behavior when the value of the volume fraction deviates from zero.

In order to study the performance of a machine learning network in the prediction of percolation effect, we first use the previously generated 10,000 training samples on Figure (13) and generate a different 1500 testing samples. Two layers of CNN with maxpooling construct the convolution part of the neural network as shown in Table (4). For both CNN and pooling layers, $4 \times 4$ kernels provide the filters along with (3,3) and (2,2) strides, respectively. The first convolution layer contains 32 filters and the second convolution layer has 64 filters. To double check the effect of the filter size, we multiplied the number of filters by 2 and 4 and did not find a noticeable change in the network performance except an increase in the training time. After the convolution layers locates three layers of affine layers, each with the size of 2. Leaky ReLU activation function enforces the nonlinearity at the end of each CNN and affine layer. To improve the training, after the first CNN layer a dropout system randomly ignores 50% of the nodes during each training epoch. Like other cases, Adams algorithm optimizes the network parameters by minimizing the absolute value error of the difference between the loss function i.e. network predictions and the DNS results. The learning rate of $0.0002$ is decreased 2% after each 500 training steps. A feeding batch size of 5, applies stochastic gradient technique for the training. With a total of 100,000 training epochs, the final error over the training set becomes 18.3%. The noticeable error of ML predictions is because the training occurs on a large input size and a diverse range of geometrical features. However, ML can provide the overall trend of the percolation phenomena and reduces the regression error from 30.4% to 18.5%.

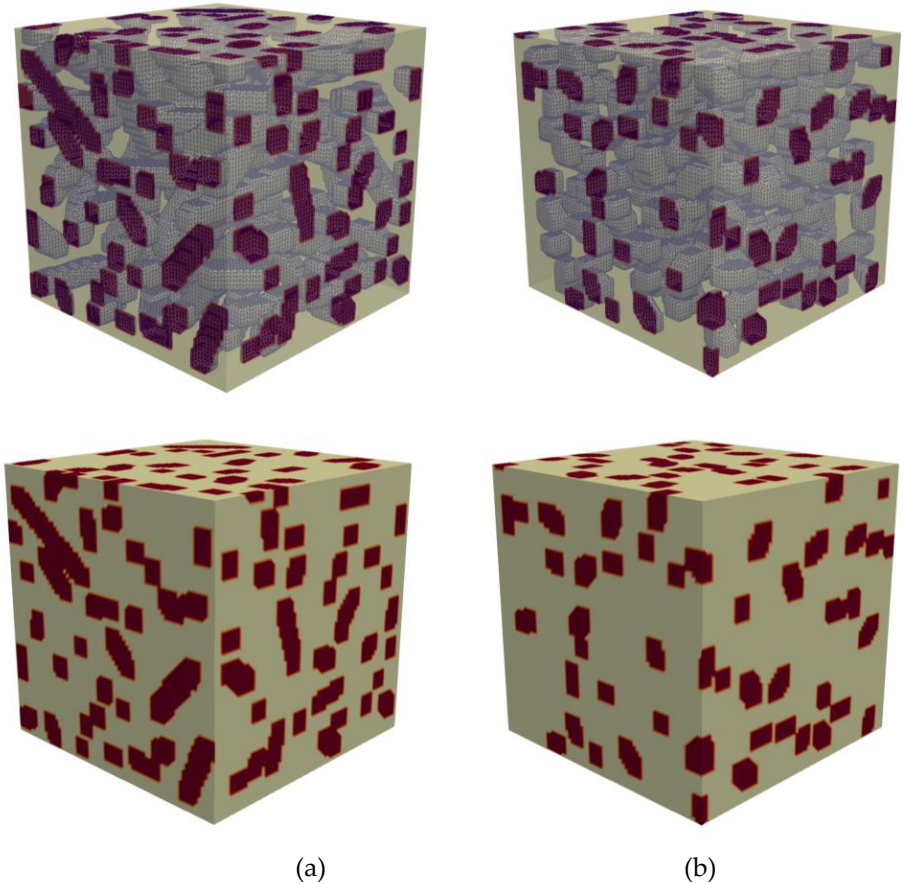

(a) (b)

*Figure 12- Generated geometries for random placement of rods and cubes. (a) The domain containing rods with $vf = 0.21$ and $k_{eff} = 4.21$ (b) The domain containing cubes with $vf = 0.21$ and $k_{eff} = 2.64$.*

Figure (14) compares ML predictions with simulation results for $vf \sim 0.28$ i.e. $0.27 < vf < 0.29$. ML can recognize the type of particles and can make predictions within the range of DNS. Also, ML predictions have better accuracy for cubic particles than for elongated particles because of high disparity on DNS results of elongated particles. With the increase of the volume fraction, the effective thermal conductivity depends more on the geometry and the shape of the embedded material. Moreover, the value of the effective thermal conductivity increases dramatically.

Table 4- The network shape to predict the percolation effect

|  | Input | CNN1 | Max pool 1 | CNN2 | Max pool 2 | 3 affine layers |
|---|---|---|---|---|---|---|
| **Size** | $72 \times 72 \times 72$ | $216 \times 192 \times 32$ | $108 \times 96 \times 32$ | $32 \times 36 \times 64$ | $18 \times 16 \times 64$ | 2 |

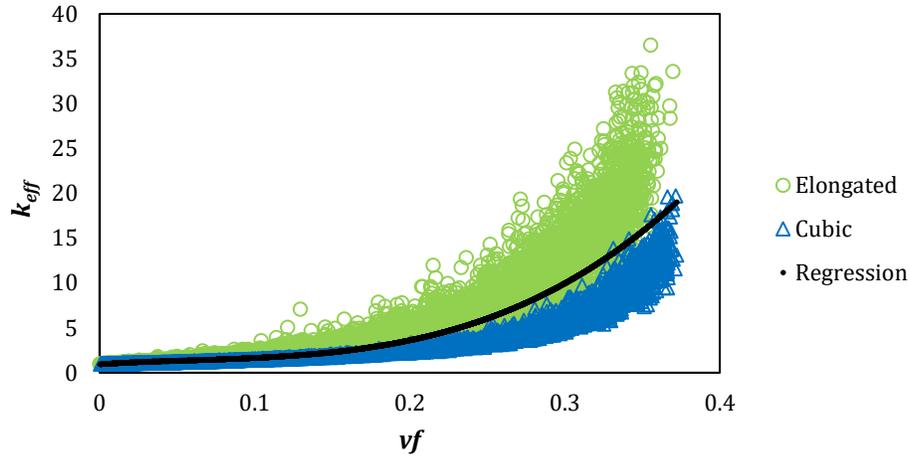

Figure 13- Variation of effective thermal conductivity with the variation of volume fraction for the number of 5000 random non-overlap rods with size $30 \times 5 \times 5$ and 5000 random non-overlap cubes with the size $5 \times 5 \times 5$ in a $72 \times 72 \times 72$ domain. A regression line shows the how in average effective thermal conductivity changes with volume fraction.

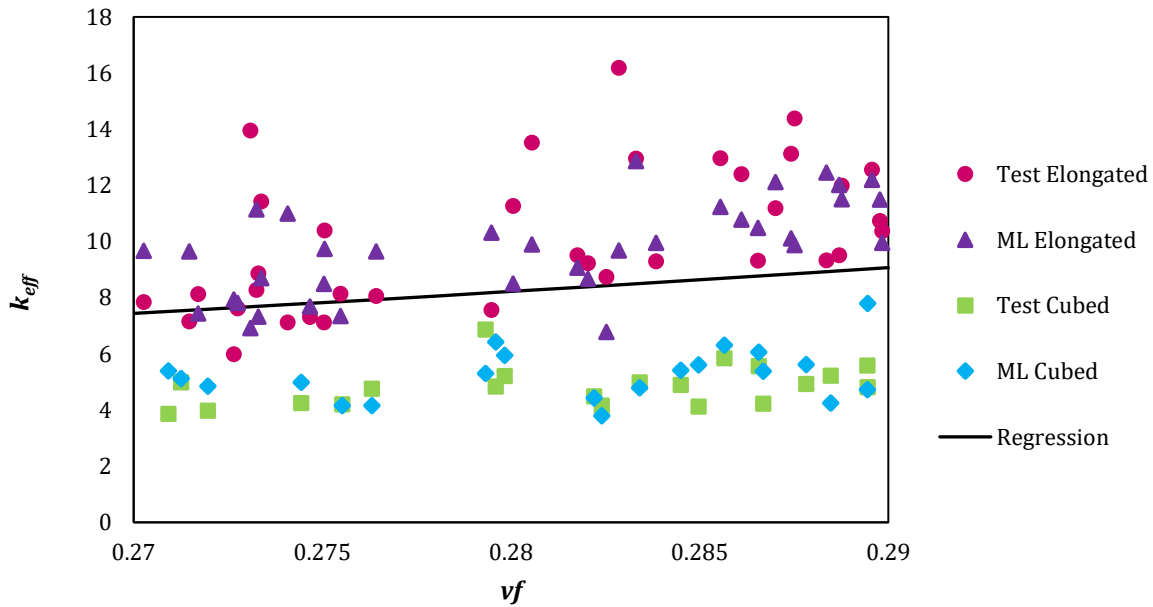

Figure 14- Comparison of the predictions from the simulation and from machine leaning. ML can distinct between the two types of particles.

**Conclusion**

A new approach is presented here to find the thermal properties of lithium-ion batteries based on actual images. This approach constructs 3D geometries from micro-images of electrodes and uses representative elemental volumes (REVs) from random places in the constructed images. With random values of conductivity, each REV is solved with a DNS solver to find its effective conductivity. A machine learning algorithm based on convolutional neural nets is set up and trained to take the input geometry as the matrix of conductivities to result in the effective conductivity of the REV. Two different networks were constructed and compared. The first network uses a coarser grid based on the average values of conductivities as the input. The other one, uses different numbers of cross sections from the 3D geometry. Finally, the applicability of using a neural net to predict percolation was studies.

The results show that the implementation of an average-based coarse grid to generate the input to the network leads to small changes in the network error. This is because the increase of the error due to updating the input from the coarse grid is compensated with a better trained network, which has a smaller input size. On the other hand, use of coarser grids can decrease the computational time by about three orders of magnitude. The method successfully combines the images from several electrodes, which are different based on the constituents' material and constructive geometries. It predicts the effective thermal conductivity with low network error. The error in the network grows with the increase of conductivity ratio because of higher sensitivity to the network training error. However, use of geometry instead of volume fraction and material conductivity has about 40% less error. The investigation of training a network based on different cross section shows that even though the increase of the number of the cross sections increases the accuracy of the network training, the changes are not that significant. Moreover, the results of the averaging technique are more accurate. Finally, we showed that a convolutional neural net can show the percolation trend and have significantly better predictions than regression analysis. The error in predicting the elongated particles is higher due to the greater disparity in the data due to the wider range of distribution.